\documentclass[twocolumn,appendixfloats,tighten]{aastex6}
\usepackage{newtxtext,newtxmath}
\usepackage[T1]{fontenc}
\usepackage{ae,aecompl}

\usepackage{amsmath}	
\usepackage{amssymb}	

\usepackage{ctable}
\usepackage{url}
\usepackage{xspace}
\usepackage[normalem]{ulem}
\usepackage{hyperref}
\usepackage[all]{hypcap}
\usepackage{graphicx}

\def\msun{{\rm\,M_\odot}}

\def\msun{{\rm\,M_\odot}}

\newcommand{\kms}{\, {\rm km\, s}^{-1}}

\newcommand{\be}{\begin{equation}}
\newcommand{\ee}{\end{equation}}

\newcommand{\Rsun}{ R_{\odot}}
\newcommand{\rsun}{ R_{\odot}}

\newcommand{\mpy}{M_{\odot}/{\rm yr}}
\def\h2{${\rm\,H_2}$}

\usepackage{color}

\begin{document}
\title{LB-1 is inconsistent with the X-ray  source population and pulsar-black hole binary searches  in the Milky Way}
\author{Mohammadtaher Safarzadeh\altaffilmark{1}, Enrico Ramirez-Ruiz\altaffilmark{1,2}, Charles Kilpatrick\altaffilmark{1}}
\altaffiltext{1}{Department of Astronomy and Astrophysics, University of California, Santa Cruz, CA 95064, USA,
\href{mailto:msafarzadeh@cfa.harvard.edu}{msafarza@ucsc.edu}}
\altaffiltext{2}{Niels Bohr Institute, University of Copenhagen, Blegdamsvej 17, 2100 Copenhagen, Denmark}

\begin{abstract}
If confirmed, a wide binary system of 70 $\msun$ black hole (BH) and an 8 $\msun$ main sequence star (LB-1) is observed to reside in the Milky Way (MW). 
While we remain agnostic about the nature of LB-1, we show that long term evolution of an 8 $\msun$ star around a BH with mass between $5-70\msun$ makes them visible as ultra-luminous X-ray (ULX) sources in the sky.
Given the expected  ULX phase lifetime  ($\approx0.1$ Myr) and their lack of detection in the MW, we conclude that the frequency of an 8-20 $\msun$ star to be in binary around a stellar mass BH should be less ($f<4\times10^{-3}$). 
This is in tension with Liu et al. (2019) claimed detection frequency of LB-1 like system around 8-20$\msun$ stars ($f\approx3\times10^{-2}$).
Moreover, the 8 $\msun$ star is likely to end as a neutron star (NS) born with a very small kick from an electron capture supernova (ECSN), leaving behind  a wide NS-BH binary. 
So far less than 1\% of all the detectable pulsars in the MW are mapped and there has been no detection of any pulsars in binary systems around BHs 
which sets an upper bound of about 100 possible pulsar-BH systems in the MW. 
We show if the NS is born from ECSN, a frequency upper limit of ($f\approx10^{-3}$) for stars with masses $\approx 8-20\msun$ in the MW to have a BH companion. 
The rate discrepancy will further increase as more pulsars are mapped in the MW, yet  these searches would  not be able to rule out the Liu et al. detection frequency
if  NSs are instead born in core collapse SNe with the commonly inferred high kick velocities.
\end{abstract}

\section{Introduction}
If confirmed, the formation of LB-1, the recently discovered 8 $\msun$ blue star orbiting a 70 $\msun$ BH in a low eccentricity orbit ($e\approx0.03$) and semi-major axis of about 300 $\Rsun$ \citep{Liu:2019ch} lacks a convincing explanation. 
In short there are two issues with this system: (i) the mass of the BH is above pair-instability pulsation supernovae (PPSN)  
for metal rich stars (Population I/II), although 70 Msun BH is allowed (at
the extreme) for metal-free stars \citep[Population III; ][]{Woosley:2017dj} while the stellar companion for this BH in LB-1 is at solar metallicity.
(ii) The 8 $\msun$ star is orbiting this massive BH at a distance of about 300 $\rsun$ which is smaller than what is expected to be the radius of the progenitor star of the 70 $\msun$ BH when
it expanded into its giant phase. Some of the stellar evolution models with reduced mass loss rates appear to be consistent with the formation of black holes up to 70 $\msun$ at high metallicity, but unable to explain how a binary star system like LB-1 could have formed without invoking some exotic scenarios \citep{Belczynski:2019uz,Tanikawa:2019tw}.

Attempts to explain LB-1 by reducing the wind mass loss efficiency in metal rich stars by a factor of about 5 have not been successful. Moreover, it has been shown that the 70 $\msun$ BH can not be a binary of two less massive BHs given the H-$\alpha$ line emission profile of this system \citep{Shen:2019vo}.
Also we note that follow up analysis have challenged the results of \citet{Liu:2019ch} indicating that LB-1 does not host a 70 $\msun$ BH \citep{AbdulMasih:2019vp,ElBadry:2019vd,Eldridge:2019wb}.

In this \emph{Letter}, we remain agnostic as to whether LB-1 hosts a 70 $\msun$ BH or a smaller BH companion. However, we ask if a wide binary system as claimed in \citet{Liu:2019ch} could exist in the Milky Way and slip detection. 
In \S2 we estimate the expected number of LB-1-like systems given the formation rate of 8-20 $\msun$ main sequence stars. 
In \S3 we present forward modeling of LB-1 and show that such systems enter a short 
ultra-luminous X-ray (ULX) phase and discuss how the lack of such objects on the sky constrains the possible number of LB-1-like systems in the Galaxy. 
In \S4 we estimate the expected number of pulsar-BH systems in the MW. 
In \S5 we discuss the survival probability of LB-1 like systems as pulsar-BH binaries given the natal kicks of the newly born NSs, and how this puts an upper limit on the total number of LB-1 like systems in the MW.
In \S6 we discuss caveats related to our work. 

\section{Expected number of LB-1 like systems in the MW}
The end life of LB-1 is a wide binary BH-NS system if the system survives the natal kick at the formation of the NS. 
We denote the formation rate of 8-20 $\msun$ stars by $R_8$.
LB-1 is found after a spectroscopic follow up on 3000 targets in K2-0 field of Kepler for stars brighter than 14 mag, which we  will  later use for comparison. 
We assume that a fraction ($f$) of all the  
stars with mass between 8-20$\msun$ are in an LB-1 like configuration.
Therefore, the formation rate of 8-20 $\msun$ stars having a massive BH companion is 
\be
R_{\rm 8,bh}=f\times R_8.
\ee
Given that the lifetime of $8 \msun$ stars is approximately 50 Myr \citep{Cummings:2018cz}, the mean age of stars with mass between 8-20 $\msun$ is about 20 Myr. Therefore, the expected number of such binary systems in the MW is 
\be
N_{\rm LB-1}\approx R_{\rm 8,bh}\times 20 {\rm Myr}\approx 4\times10^5f.
 \label{eq2}
\ee
As the end life of stars with this mass is likely a NS, we have assumed $R_8\approx0.02$ yr$^{-1}$ following \citet{Licquia:2015eb} where they assume a Kroupa IMF, which gives  $2\times10^8$ NSs formed in the MW over 10 Gyr.

The limiting magnitude of \citet{Liu:2019ch} target is $m=14$. Assuming the disk extends to $\approx14$ kpc \citep{Minniti:2011jh}, and sun is at about 8 kpc from the center \citep{Eisenhauer:2003iu}, by looking in the direction of Galactic anti-center, the distance range in \citet{Liu:2019ch} sample is from 0 to 6 kpc. 
The limiting magnitude corresponds to an absolute magnitude at each distance according to $M_{\rm abs}=14-5\log(d)+5$ where $d$ is distance 
to the source in parsecs. We can read off the corresponding luminosity of the source in solar units from $M_{\rm abs}$ at each distance by $L=10^{.04(M_{\rm abs,\odot}-M_{\rm abs})}$, where 
we take $M_{\rm abs,\odot}=4.83$.
The luminosity is then converted to mass based on mass-luminosity relation of zero-age main sequence stars \citep{Tout:1996dq}. 
This allows a minimum mass $m_{\rm min}$  observable in a 14 magnitude limited survey as a function of distance to be derived. 
The $m_{\rm min}$ we compute is about 1, 2, and 3 $\msun$ at 1, 3, and 6 kpc from the sun, respectively. 
At each distance, we compute the fraction of stars with masses between $8-20\msun$ over the stars with mass between $m_{\rm min}-150\msun$ assuming a Kroupa-type initial mass function $\zeta(m)\propto m^{-2.35}$. 
We weight the resulting fraction by the lifetime of a star with mass $m_{\rm min}$ relative to the lifetime of an 8 $\msun$ star on the main-sequence \citep[$t_{\rm ms}\propto m^{-2.92}$;][]{Demircan:1991ee}. We finally derive the total fraction weighted by the volume at each distance shell ($\propto d^2$) and  conclude that the expected total number of $8-20\msun$ stars in \citep{Liu:2019ch} is about 1\% of their sample. 
Therefore, the detection rate of LB-1 like systems around 8-20 $\msun$ stars in \citet{Liu:2019ch} is about 1 in 30 ($f\approx3\times10^{-2}$). 
We note that this is a lower limit in that we have not yet taken into account the overall drop in stellar density with Galactic radii which could increase the detection frequency above 1 in 30. 
This means that according to \citet{Liu:2019ch} the total number of LB-1 like systems in the Galaxy is more than about $1.4\times10^4$ (see Eq. \ref{eq2}).

To account for the dust extinction we perform the following analysis: The limiting magnitude of \citet{Liu:2019ch} target is $m_{\rm Kepler}=14$~mag in the Kepler bandpass and all stars were observed in a narrow ($\approx$15$\times$15~deg$^{2}$) centered at Galactic coordinates $l=191.64^{\circ}$, $b=5.85^{\circ}$. We estimate the corresponding initial mass limit by assuming that the Milky Way is represented by a disk with an exponential density profile in radius and height \citep{Kalberla08} with atomic hydrogen density:

\begin{equation}
    n_{H} = n_{H,0} e^{-\frac{r-R_{0}}{R_{n}} - \frac{z}{H}}
\end{equation}

where $n_{H,0}$=0.9~cm$^{-3}$, $R_{0}$=8~kpc, $R_{n}$=3.5~kpc, $H$=300~pc, $r=\sqrt{x^{2}+y^{2}}$ is the Galactocentric radius, and $z$ is the disk height.  We inject stars with a known initial mass into the Milky Way disk with their locations $(x,y,z)$ in kpc toward the Galactic coordinates for K2-0 above and where the Sun is located at $(x,y,z)=(8,0,0)$, that is with $(x,y,z)=(-d \cos(b) \cos(l) + R_{0}, -d \cos(b) \sin(l), d \sin(b))$ and with $d\in[0,6]$~kpc, $l\in[184.1,199.2]$~deg, $b\in[-1.7,13.4]$~deg.  As star formation traces hydrogen density, we assume that the number density of stars as a function of Galactocentric radius and disk height is $n^{*} \propto n_{H}$ and inject $10^{6}$ stars following a probability density function equal to the normalized atomic hydrogen density at their location.  We then calculate the column of atomic hydrogen along the line of sight to the Sun and convert that quantity to a $V$-band extinction following \citet{Guver09} and then to extinction in the Kepler bandpass using {\tt pysynphot} and the Kepler filter transmission function ($A_{\mathrm{Kepler}}\approx0.83 A_{V}$).  The average line-of-sight extinction in the K2 bandpass is $A_{\mathrm{Kepler}}=1.29$~mag.

\begin{figure}
    \centering
    \includegraphics[width=0.49\textwidth]{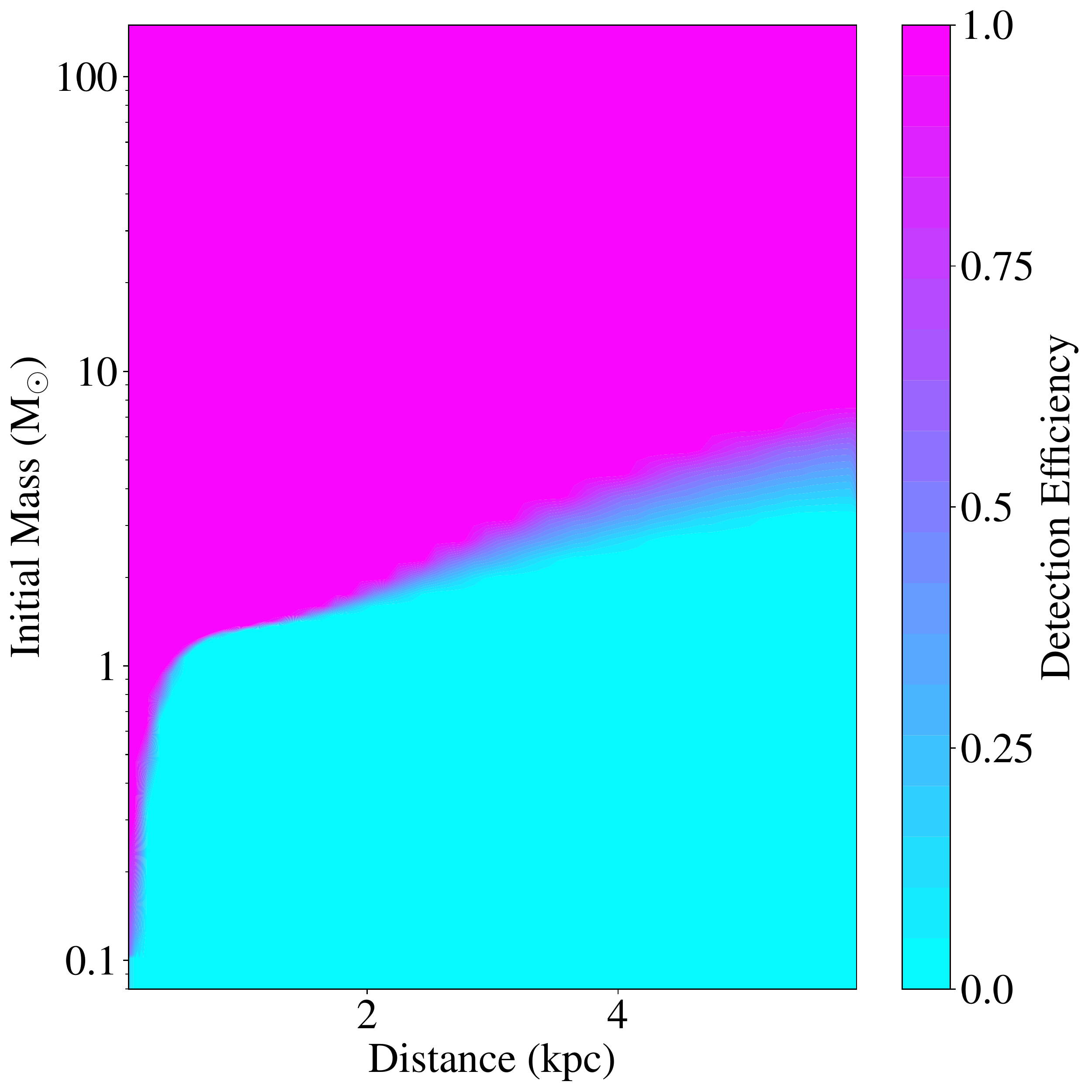}
    \caption{Detection efficiency of stars with a given initial mass and distance from the Sun assuming that all stars are detected in the Kepler bandpass with  $m_{\mathrm{Kepler}}<14$~mag.  We inject stars into the Milky Way disk and account for dust extinction following the procedure described in \S2.}
    \label{fig:efficiency}
\end{figure}

We determine the absolute magnitude of stars with a given initial mass close to the terminal age main sequence (TAMS) by analyzing Mesa Isochrone \& Stellar Tracks (MIST) single-star evolutionary tracks \citep{Choi16,Choi17} where the star has reached 10\% hydrogen mass fraction in its core.  Assuming that stars in the field are well represented by these tracks, we look up the Kepler in-band absolute magnitude at this stage in the star's evolution, which is calculated directly by MIST.  Finally, we calculate the recovery fraction of stars with a given initial mass and in different mass bins by determining the fraction of stars detected with $ m_{\mathrm{Kepler}}<14$~mag along all lines of sight.  The fraction of stars we recover averaged along all lines of sight as a function of initial mass and distance ($E(m,d)$) is shown in \autoref{fig:efficiency}.

For reference, the initial mass at which 50\% of stars are recovered is about 1.3, 2.3, and 4.5 $\msun$ at 1, 3, and 6 kpc from the Sun, respectively. At large separations from the Sun, these mass estimates are strongly affected by dust extinction as represented by the atomic hydrogen column density along different lines of sight.

Using these detection efficiencies, we then ask what is the intrinsic fraction of stars with initial masses from 8--20$\msun$ relative to stars detectable in the magnitude-limited simulation above?  Assuming a Kroupa-type initial mass function with $\zeta(m)\propto m^{-2.35}$ for $0.5~\msun < m$ and $m^{-1.3}$ for $0.08~\msun < m < 0.5~\msun$, we weight the resulting mass bins by the lifetime of a star with a particular mass relative to the lifetime of a 8~$\msun$ star on the main sequence \citep[$t_{\rm ms}\propto m^{-2.92}$;][]{Demircan:1991ee}.  Integrating for Heliocentric distances $<6$~kpc and initial masses $0.08~\msun < m < 150~\msun$, we calculate the following fraction,

\begin{equation}
    \eta = \frac{\int_{V} \int_{8}^{20} \zeta(m) t_{\rm ms} n^{*}(d,\Omega) E(m,d, \Omega)~\mathrm{d}m~\mathrm{d}V}{\int_{V} \int_{0.08}^{150} \zeta(m) t_{\rm ms} n^{*}(d,\Omega) E(m,d,\Omega)~\mathrm{d}m~\mathrm{d}V}.
\end{equation}

where $n^{*}(d,\Omega)$ is analogous to the normalized atomic hydrogen density used as a proxy for stellar density above, but we account for different likelihoods of finding a star along different lines of sight $\Omega$ (covering all solid angle viewed with respect to the Sun, rather than averaged as in \autoref{fig:efficiency}) and at different Heliocentric distances $d$.  Similarly, we account for different efficiencies with respect to line of sight in $E(m,d,\Omega)$.  We then integrate over all mass bins $dm$ and volume $V$ within a Heliocentric distance $d=6$~kpc, with volume elements $dV$ subdivided into Heliocentric distance elements $dd$ and solid angle elements $d\Omega$.  Under these assumptions, we estimate that the expected total number of 8--20~$\msun$ stars in \cite{Liu:2019ch} is about $\eta=0.95$\% of their sample.
Therefore, the detection rate of LB-1 like systems around 8--20 $\msun$ stars in \citet{Liu:2019ch} is approximately 1 in 29 (i.e., 1 in 0.95\% of 3000, $f=3\times10^{-2}$). 
This means that according to \citet{Liu:2019ch} the total number of LB-1 like systems in the Galaxy is more than about $14,000$ (see Eq. \ref{eq2}).
 
 \section{forward evolution modeling of LB-1 like systems}
 \label{sec:forward_modeling}
We study the forward evolution of LB-1 based on models discussed in \citet{Mondal:2019ha} where they are based on StarTrack population synthesis analysis \citep{Belczynski:2008kt}.
 
We begin the evolution of our binary system at time $t=37.446$ Myr when the initially
$8.0\msun$ star finishes its main sequence evolution. At this moment, 
the binary consists of a $M_2=7.9\msun$ optical component orbiting a BH with mass
$M_1=68.0\msun$; orbital semi-major axis is $a=327.6\rsun$ and eccentricity
is $e=0.0$ (orbital period $P=78.9$d). The optical component at this point
($R_2=8.1\rsun$) is well within its Roche lobe radius ($R_{l,2}=70.8\rsun$).

The optical star is subject to rapid radial expansion on the Hertzsprung gap and
at time $t=37.54$ it overflows its Roche lobe ($R_2=71.0\rsun$), initiating a
stable mass transfer phase onto the BH. At this point, the optical star $M_2=7.8\msun$
has a He core mass of $M_{2,\rm He}=1.6\msun$.

The Roche lobe overflow (RLOF) mass transfer rate is very high $dM/dt_{\rm rlof}=1.6 \times 10^{-4}\mpy$, 
and in particular it significantly exceeds the Eddington limit for a $68\msun$ BH: 
$dM/dt_{\rm Edd}: 1.8 \times 10^{-6} \mpy$. Note that we only allow a small
fraction of this mass to be accreted onto the BH: $dM/dt_{\rm acc}: 7.5 \times 
10^{-7} \mpy$, while the rest of the mass is assumed to be ejected from the system with angular momentum characteristic of the BH component \citep[as in model presented by][]{Mondal:2019ha}.
Note that this marks the onset of a very bright ULX phase, as the isotropic X-ray
luminosity of the system is $L_{\rm x,iso}=5.5 \times 10^{40}$ erg/s, and
if it is beamed it may reach  $L_{x,\rm beamed}=5.8 \times 10^{42}$ erg/s \citep{Mondal:2019ha}. 

RLOF ends at $t=37.60$ Myr, when the H-rich envelope of the optical component is 
 significantly depleted ($M_2=2.7\msun$ with $M_{2,\rm He}=1.6\msun$). At this stage,  the star has evolved to core-Helium burning and starts contracting instead of 
expanding. At this point the mass transfer slows down to $dM/dt_{\rm rlof}=6.9 
\times 10^{-5}\mpy$, but it is still high enough that the system continues to be a very
bright ULX: $L_{\rm x,iso}=4.7 \times 10^{40}$ erg/s and $L_{x,\rm beamed}=9.7 \times 
10^{41}$ erg/s. At the end of this evolutionary phase, the BH has not gained  significant mass: $M_1=68.04\msun$ as most of the transferred mass ($5.1\msun$) was ejected from 
the system. The mass loss from binary leads to a significant increase in the
binary separation: $a=2575\rsun$ ($P=1801$d).  

\begin{figure}
  \includegraphics[width=0.9\columnwidth]{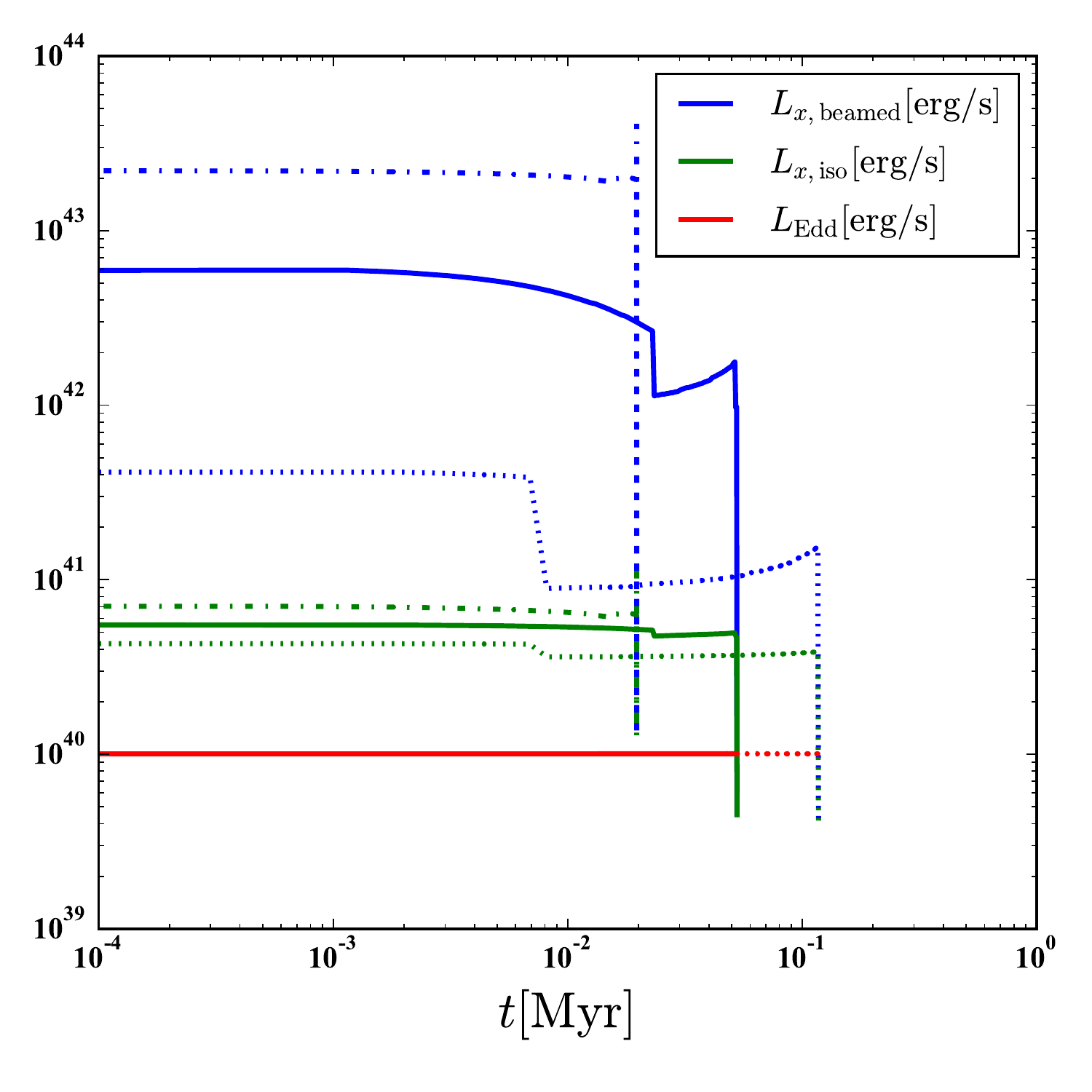}
 \caption{The X-ray luminosity evolution of a wide binary consist of a massive BH of mass about 70 $\msun$ and a companion star with mass 8 $\msun$ (solid lines), 
 6 $\msun$ (dotted lines), and 12 $\msun$ (dot-dashed lines) from the onset of RLOF considered to be $t=0$. For the 6 $\msun$ case, the system enters RLOF phase twice but only the first phase is shown in this Figure. 
 The isotropic, beamed, and Eddington luminosities for each case is shown in blue, green, and red respectively. Both the isotropic and beamed X-ray luminosities of all the three systems exceed the Eddington luminosity of the accreting BH by a large factor. 
 Therefore, irrespective of the precise mass of the companion, such wide binaries are observable as a ULXs on the sky.}
 \label{fig:evolv}
\end{figure}

At $t=41.8$ Myr, the optical star has lost the remainder of its H-rich envelope via stellar winds and it has become a naked helium star with mass $M_2=2.12\msun$. At $t=42.48$ Myr, the evolved helium star with mass $M_2=2.04\msun$ and CO
core of mass $M_{2,\rm He}=1.38\msun$ is assumed to undergo an electron-capture supernova (ECSN), forming a low mass NS ($M_2=1.26\msun$). An explosion happens on the circular
and very wide orbit: $a=2663\rsun$ (orbit has expanded due to wind mass loss from the optical star). 
Depending on the adopted natal kick \citep[if any in case of an ECSN;][]{Gessner:2018by} and mass loss ($\approx0.7\msun$; Blaauw kick) the
system either survives as BH-NS binary or is disrupted forming two single
compact objects. We investigate the impact of natal kicks in details in the next section. The evolution described above is shown with solid lines in Figure \ref{fig:evolv}. 

We have carried out two more simulations. In the first, we evolve a 6 $\msun$+68 $\msun$ BH. 
This system enters a stable RLOF and becomes ULX between which lasts 0.117 Myr (from 68.819 to 68.936 Myr). 
Then the star detaches (contraction during core Helium burning). The star re-expands during AGB evolution entering a second phase of RLOF that lasts 0.197 Myr (from 78.216 to 78.413 Myr). 
At $t$=78.58 Myr the system ends as a 1.39 $\msun$ carbon-oxygen white dwarf (WD) and a 68.51 $\msun$ BH at 2873 $\rsun$. The first RLOF evolution of this system is shown with dotted lines in Figure \ref{fig:evolv}. 
In the second simulation, we evolve a 12 $\msun$+68 $\msun$ BH starting with a=333 $\rsun$.  
This system enters a stable RLOF and becomes ULX between 17.928-17.941 Myr. This is much shorter than the timescale in which the 6+68 system lasts as ULX. 
After RLOF the system is 68.01 $\msun$ BH + 2.99 $\msun$ star with a He core mass of 2.7$\msun$ at a separation of 4374 $\rsun$.
At t=20.85 Myr a supernovae explosion takes place in a system composed of a 2.5 $\msun$ ($M_{\rm CO}=1.63 \msun$) at 4459 $\rsun$. 
This leaves a 1.11 $\msun$ NS in a core collapse supernova (CCSN). This above evolution is shown with dot-dashed lines in Figure \ref{fig:evolv}.

Therefore, regardless of the exact mass of the secondary, such a wide binary system is expected to enter a ULX phase, with similar lifetimes, that can not be missed in the X-ray sky \citep{Grimm2002}.
As before, the expected number of such ULXs is given by :
\be
N_{\rm ulx}\approx f_{\rm b,ulx} R_{\rm 8,bh} \times t_{\rm ulx} \approx 2\times10^3f,
\ee
where $t_{\rm ulx}$ denotes the lifetime of the ULX phase  (assumed to be 0.1 Myr). 
The accretion disk in these systems are Eddington limited in all directions and not beamed, and therefore we use a beaming factor of $f_{\rm b,ulx}=1$. 
The rejection confidence level (CL) in the absence of any active ULX in the sky is given by ${\rm CL}=N_{\rm ulx}/\sqrt{N_{\rm ulx}}$, 
from which values of $f\gtrsim4\times10^{-3}$ are thus ruled out at 99\% CL given the non-detection of such systems \citep{Grimm2002}. 
This is in tension with the detection rate of \citet{Liu:2019ch} which we estimate to be $f\approx3\times 10^{-2}$. 
We note that if the optical star has a mass less than 8 $\msun$, our calculations above needs to be revised accordingly.

\section{Could LB-1 host a less massive BH?}
Recent re-analysis of the LB-1 system suggest that the BH could have a mass between $5-20\msun$ \citep{ElBadry:2019vd}.
We performed our simulations of a wide binary consisting of an 8 $\msun$ star with a 5 and 10 $\msun$ BHs and found that in both these cases the system becomes a ULX source for a similar lifetime of about 0.1 Myr. 
Therefore, our results in section \ref{sec:forward_modeling} holds the same for all the stellar mass BHs as a companion for an 8 $\msun$ star in wide binary system. 
We conclude that for LB-1 to have a BH companion of any mass is in tension  
with the lack of detection of ULXs in the sky as long as optical component is a massive B star ($>6 \msun$). 
 
\section{Expected number of wide pulsar-BH systems in the MW}
We assume that the formation rate of neutron stars in the MW is the same as formation rate of 8-20$\msun$ stars
\be
R_{\rm ns}\approx R_{8}.
\ee 
Of these NSs, a fraction ($f_{\rm bh}$) will be born in binary systems with BH companion, and of those, a fraction ($f_{\rm wide}$) will be in wide binary systems. 
Therefore, the formation rate of NSs in wide BH-NS binaries is:
\be
R_{\rm ns}^{\rm bh,wide}\approx f_{\rm bh} f_{\rm wide}R_{\rm ns}.
\ee 
The total expected number of such NSs detectable as normal pulsars is :
\be
N_{\rm plsr}^{\rm bh,wide}\approx R_{\rm ns}^{\rm bh,wide} t_{\rm plsr} f_{\rm b,plsr} \approx 4\times10^5 f_{\rm bh} f_{\rm wide} 
\ee
where we have taken $t_{\rm plsr}=100$ Myr as a typical age of a normal pulsar in the MW \citep{Halpern:2010kf}, and $f_{\rm b, plsr}=0.2$ as a typical beaming factor for the normal pulsars \citep{OShaughnessy:2010im}. 
The total number of pulsars agrees with the expected number of them after correcting for their beaming factor \citep{Kaspi:2010fx,Keane:2015wk}.

We take $f_{\rm bh}\approx10^{-4}$, and $f_{\rm wide}\approx0.5$ for NSs to be in wide binaries around BHs \citep{Olejak:2019wj}.
Therefore, we expect about $N_{\rm plsr}^{\rm bh,wide}\approx20$ systems in the MW, while $N_{\rm plsr}^{\rm bh,wide}=0$ is certainly possible given the current non-detection of  NS-BH systems in the Galaxy. 
We note that this statistic does not further divide the sample based on BH mass.

\section{Survival chance of LB-1 like systems as wide NS-BH binaries}

In this section, by LB-1 like system we mean a system with characteristics similar to the current state of LB-1. A wide binary system of a 8 $\msun$ star around a massive BH evolves first to a stable mass transfer through RLOF. 
At this stage the donor loses mass (about $\approx5 \msun$) which leads to the expansion of the orbit by \citep{Toonen:2016tw}:
\be
\frac{a_f}{a_i}=\left(\frac{m_{\delta,i}m_{\alpha,i}}{m_{\delta,f}m_{\alpha,f}}\right)^2,
\ee 
where the subscript $i$ and $f$ denote the pre- and post- mass transfer values for the donor and accretor (denoted by $\delta$ and $\alpha$ respectively).

After the expansion, the He core of the stripped star experiences a supernova explosion resulting in a kick imparted to the newly born NS. 
    The magnitude of this kick can be very small (order of a few $\kms$) for an ECSN explosion, or larger for a Fe core collapse explosion of a more massive He core \citep[e.g.][]{Holland2017}.
After this SN explosion, the system can remain bounded depending on the mass loss experienced during the explosion  ($\Delta M$), the total mass of the binary before the explosion ($M_{\rm tot}$), 
the magnitude of the kick ($v$), and the relative velocity of the neutron star with respect to the central mass of the system ($v_{\rm rel}$).
Averaged over random kick directions, the probability the binary remains bounded is given by \citep{Hills:1983cs,Tauris:2017cf}:

\begin{equation}
  P_{\rm bound}=\frac{1}{2}\, \bigg\{1+\left[\frac{1-2\Delta M/M_{\rm tot} -(v/v_{\rm rel})^2}{2\,(v/v_{\rm rel})}\right]\bigg\} \;.
\label{eq:bound}
\end{equation}

This functional form is bounded between zero and one. 
We assume about 5 $\msun$ is lost from a binary in RLOF phase from the 8 $\msun$ star which leads to an orbital expansion factor of about 9, after which we consider a SN kick.

We assume that natal kick of the neutron stars follows a Maxwell-Boltzmann (MB) distribution. Analyzing the proper motion of 233 pulsars \citet{Hobbs:2005be} arrive at an MB distribution with  rms 1d velocity dispersion of 265 $\kms$.
However, if the end product of the 8 $\msun$ star is a NS, likely it is not coming from a massive Fe core collapse with large kicks, but rather from ECSN \citep{Nomoto:1984bb} with small kicks \citep{Suwa:2015bs,Janka:2016df}.

\begin{figure}
\includegraphics[width=0.9\columnwidth]{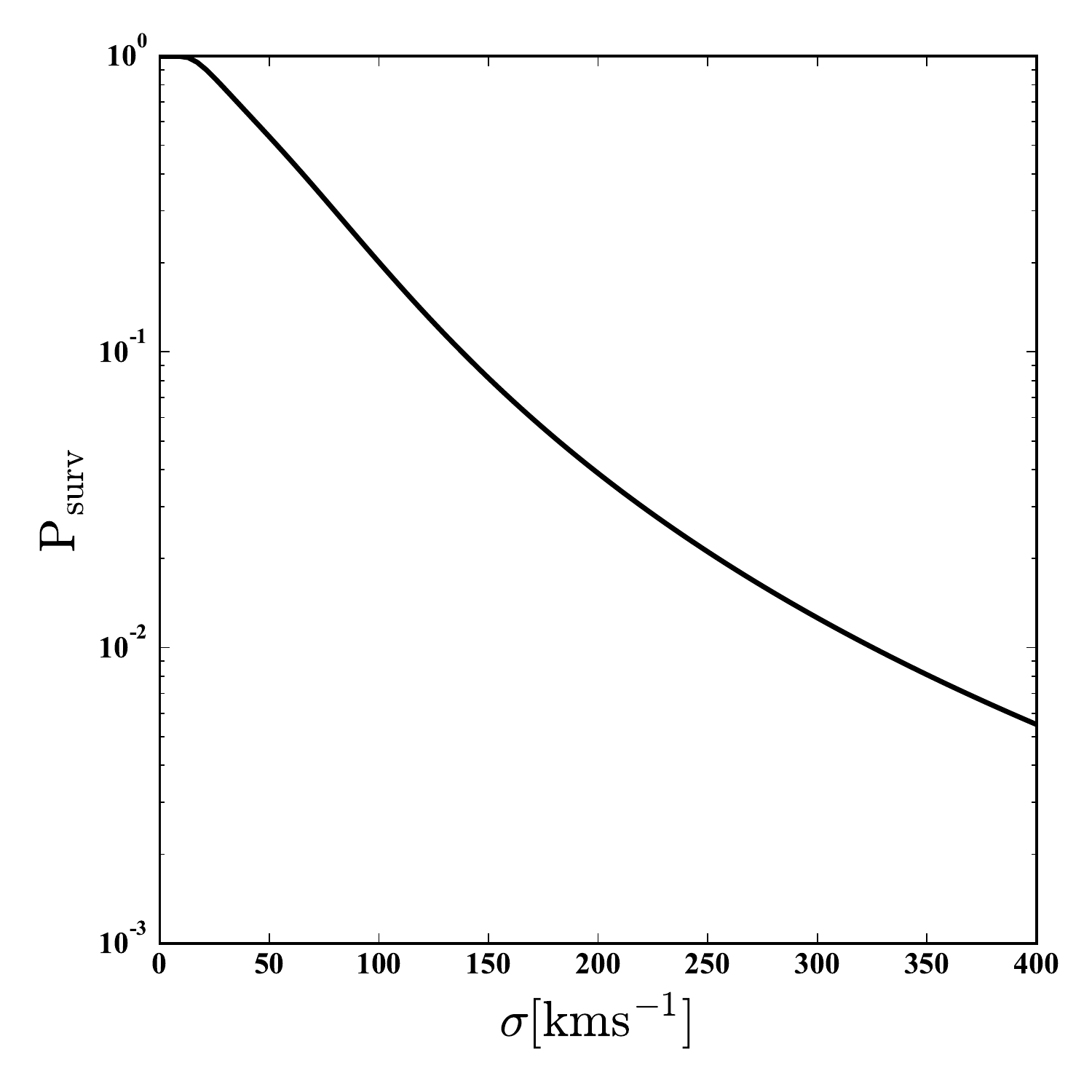}
\caption{The overall survival chance of an 8 $\msun$ star orbiting a 70 $\msun$ BH as a function of the 1d velocity dispersion of the MB distribution describing the kick magnitude the NS receives at birth.
We have assumed the binary period prior RLOF expansion is 80 days in this calculation.}
\label{fig:p_all}
\end{figure}

The overall survival probability of such systems can be computed as
\be
P_{\rm surv}= \int P_{\rm bound}(v) P (v) dv.
\ee
Here we assume $P(v)$ follows an MB distribution.
Figure \ref{fig:p_all} shows the effect of $\sigma$ on the survival chance if we assume that binary period prior RLOF expansion was 80 days.
For example with $\sigma\approx130\kms$ we expect 10\% of such systems to survive. 

 \begin{figure}
 \includegraphics[width=0.9\columnwidth]{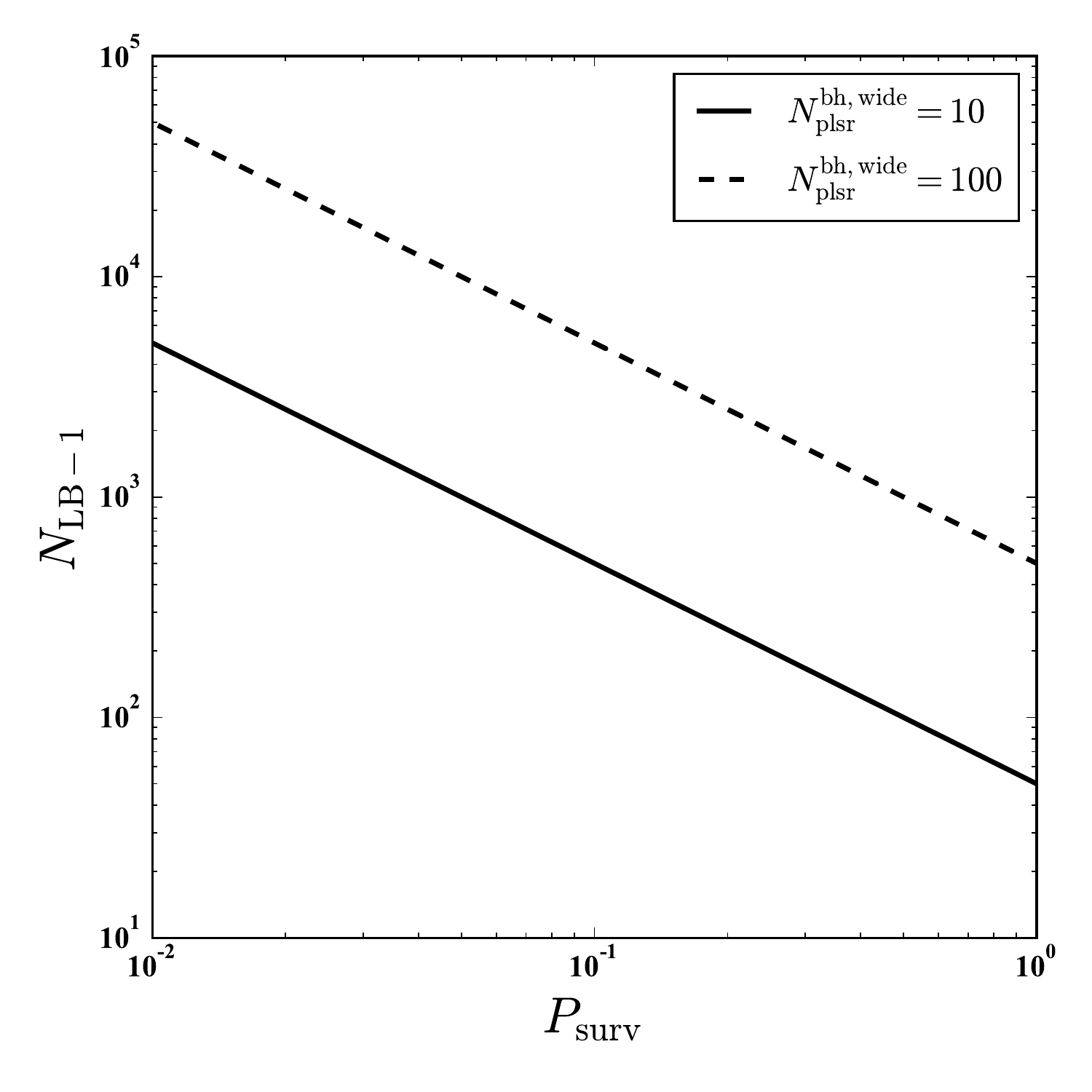}
 \caption{The upper limit on $N_{\rm LB-1}$ as a function of $P_{\rm surv}$ assuming $N_{\rm plsr}^{\rm bh,wide}=10\,(100)$ shown with solid (dashed) line. 
If NS natal kick is less than a few $\kms$ which is expected to be in the case of EC SNa, LB-1 like systems would end as wide pulsar-BH binaries. 
In this case, if the entire pulsars in the MW are mapped and less than 100 pulsar-BH system is found, no more than 500 LB-1 like system could exist in the MW. 
This number scales linearly with the number of future pulsar-BH binary detections.}
 \label{fig:confidence}
\end{figure}

The maximum number of detectable radio pulsars in wide binaries around BHs as end life of LB-1 like systems is given by:
\be
N_{\rm plsr}^{\rm bh,wide}\gtrsim P_{\rm surv} N_{\rm LB-1}f_{\rm b,plsr},
\label{eq:10}
\ee
where $P_{\rm surv}$ is the survival chance of LB-1 like systems to end as wide NS-BH binaries and not get disrupted by the natal kick.
This is an upper limit for the product of $P_{\rm surv} N_{\rm LB-1}$ since there might exist alternative pathways to make wide pulsar-binary black hole in the nature than endpoint of LB-1 like systems. 

Figure \ref{fig:confidence} shows the upper limit on $N_{\rm LB-1}$ as a function of $P_{\rm surv}$ assuming $N_{\rm plsr}^{\rm bh,wide}=10\,(100)$ shown with solid (dashed) line.
Since such wide binaries make NSs with very small kicks, the $P_{\rm surv}$ is effectively unity for such systems. 

So far about 1\% of all the detectable pulsars in the MW have been mapped \citep{Lorimer2008,Stovall2013} and there has been no detection of any pulsars in binary systems around BHs.
This sets an upper bound of about 100 possible pulsar-BH systems in the Galaxy that are yet to be detected. 
Given Eq. \ref{eq:10} an upper limit of 500LB-1 like systems could reside in the MW if less than 100 pulsar-BH system is found when the entire pulsars in the MW are mapped. 
Therefore, stars with mass $\approx 8\msun$ in the MW could have a massive BH companion with a frequency less than $f=1.25\times10^{-3}$ (Eq. \ref{eq2}).
This is in strong tension with the detection rate of such system based on \citet{Liu:2019ch} results.
We note that this is assuming that the NS is born in ab ECSN with a very small kick. If, however, the NS is born from CCSN with high kicks such that $P_{\rm surv}=0.01$, 
the pulsar searches would not be able to rule out the detection frequency of \citet{Liu:2019ch}.

\section{Summary and Conclusions}
Recently \citet{Liu:2019ch} claimed discovery of LB-1, a wide binary system of a 70 $\msun$ BH and an 8 $\msun$ main sequence stars in the Galaxy. 
This detection has recently been challenged and no viable formation mechanism has been put forth for such a system.

In this \emph{Letter} we remain agnostic about how such systems could have formed, and instead investigate the upper bound on the number of LB-1 like systems in the Galaxy. 
We show that in their long term evolution, such systems become ULX sources, independent of the BH mass, after which they end as wide NS-BH binaries if they survive the natal kick imparted to their newly born NS. 
Given the birth rate of NSs in the MW, and the lifetime of ULX sources based on our simulations, we show that the frequency of an 8-20 $\msun$ star to be in binary around a stellar mass BH should be less 
($f<2\times10^{-3}$) which is in tension with \citet{Liu:2019ch} claimed detection frequency of LB-1 like system around 8-20$\msun$ stars ($\approx3\times10^{-2}$). 

Moreover, the 8 $\msun$ star likely ends as a NS born with  a very small kick from an ECSN which leaves the final endpoint of the system as a wide NS-BH binary. 
So far less than 1\% of all the detectable pulsars in the MW are mapped and there has been no detection of any pulsars in binary systems around BHs 
which sets an upper bound of about 100 possible pulsar-BH systems in the Galaxy. 
If the NS is born from ECSN, we show that the current null detection of pulsar-BH systems implies a frequency upper limit of ($f=5\times10^{-4}$) for stars with masses $\approx 8-20\msun$ in the MW to have a BH companion. 
The discrepancy with the \citet{Liu:2019ch} rate will  increase even further as more pulsars are mapped in the MW and are found to have no BH companions.
\\
\acknowledgements
We are thankful to the referee for their constructive comments. We are thankful to Chris Belczynski and Aleksandra Olejak for providing us the data for the ULX part of this work, and Scott Ransom, Ryan Foley, Duncan Lorimer, and Josiah Schwab for helpful discussions. MTS and ER-R thank the Heising-Simons Foundation, the Danish National Research Foundation (DNRF132) and NSF (AST-1911206 and AST-1852393) for support. 
KB acknowledges support from the Polish National Science Center (NCN) grant Maestro: 2018/30/A/ST9/00050.

\bibliographystyle{apj}
\bibliography{the_entire_lib}

\end{document}